\def\rQCED{{\rm QCED}}
\newcommand{\mathswitchr}[1]{\relax\ifmmode{\mathrm{#1}}\else$\mathrm{#1}$\fi}

\newcommand{\FYFS}{F_{\mathrm YFS}}
\documentclass{PoS}
\usepackage{graphicx}
\usepackage{amsmath}
\usepackage{amssymb}
\usepackage{epsfig}
\usepackage{verbatim}
\title{HERWIRI1.031: New Approach to Parton Shower MC's in Precision QCD Theory}

\ShortTitle{HERWIRI1.031: New Approach to Parton Shower MC's in Precision QCD Theory}

\author{\speaker{B.F.L. Ward}%
         \thanks{Work supported in part by US DOE grant DE-FG02-09ER41600.}\\
        Baylor University\\
        E-mail: \email{BFL\_Ward@baylor.edu}}

\author{S.A. Yost%
        \thanks{Work supported in part by US DOE grant DE-SC0000528 and The Citadel Foundation.}\\
       The Citadel\\
        E-mail: \email{scott.yost@citadel.edu}}

\abstract{The new IR-improved 
Dokshitzer-Gribov-Lipatov-Altarelli-Parisi-Callan-Symanzik (DGLAP-CS) kernels recently developed by one of us is implemented in
the HERWIG6.5 environment to generate a new MC, HERWIRI1.0(31),
for hadron-hadron scattering at high energies.
The comparison between the parton shower generated by the 
standard DGLAP-CS kernels and that generated by the new IR-improved 
DGLAP-CS kernels is illustrated using MC data. This is done
for some of the respective exact ${\cal O}(\alpha_s)$ corrected spectra 
using the seamless interfaces to MC@NLO while making comparisons with FNAL 
data. Some discussion of possible 
implications for LHC phenomenology is also presented.\\
\vskip0.5cm
\centerline{BU-HEPP-10-05, ~Nov., 2010}}

\FullConference{35th International Conference of High Energy Physics - ICHEP2010,\\
                 July 22-28, 2010\\
                 Paris, France}
\begin{document}
Resummed 
${\cal O}(\alpha_s^2L^n),{\cal O}(\alpha_s\alpha L^{n'}),
{\cal O}(\alpha^2 L^{n''})$ corrections for $n=0,1,2,$ $n'=0,1,2,$ $n''=1,2$, 
in the presence of parton showers, on an event-by-event basis, 
without double counting and with exact phase space are required~\cite{qedeffects1,radcor-ew1}
for precision LHC physics($1\%$ or better total theoretical precision~\cite{jadach1a}).
We present here the first step in realizing our 
new MC event generator approach to such physics
with amplitude-based QED$\otimes$QCD resummation~\cite{qced1} 
in the HERWIG6.5~\cite{herwig1} environment
by our
new parton shower MC for QCD, HERWIRI1.0(31)~\cite{irdglap3-plb-prd}. 
HERWIRI1.0(31)
already shows improvement in
comparison with the FNAL rapidity and soft $p_T$ data on single $Z$ production
as we quantify below. While the 
explicit IR cut-offs in the HERWIG6.5
environment will not be removed here, HERWIRI 
only involves integrable
distributions so that these cut-offs could be removed. 
\par\indent
We first review
our approach to resummation, which can be shown~\cite{irdglap3-plb-prd,irdglap1a} to be equivalent to those in Refs.~\cite{cattrent1,scet1}, before we turn to
a summary of the attendant new IR-improved DGLAP-CS~\cite{dglap1,cs1} theory~\cite{irdglap1a} and a description of the implementation of the new IR-improved kernels in the framework of HERWIG6.5~\cite{herwig1} to arrive at HERWIRI1.0(31). We illustrate the effects of the IR-improvement and compare with recent 
data from FNAL\footnote{From Ref.~\cite{scott1a} the current state-of-the-art theoretical precision tag on single $Z$
production at the LHC at $14$ TeV is 
$(4.91\pm0.38)\%=(2.45\pm0.73)\%(QCD+EW)\oplus 4.11\%(PDF)\oplus1.10\pm0.44\%(QCD Scale)$ and for single W$^+$(W$^-$) $5.05\pm0.58\%(5.24\pm0\%)$.}.\par\indent
In Refs.~\cite{qced1,irdglap1a}
we have derived the following expression for the 
hard cross sections in the SM $SU_{2L}\times U_1\times SU_3^c$ EW-QCD theory{\small
\begin{eqnarray}
d\hat\sigma_{\rm exp} &=& e^{\rm SUM_{IR}(QCED)}
   \sum_{{n,m}=0}^\infty\frac{1}{n!m!}\int
\frac{d^3p_2}{p_2^{\,0}}\frac{d^3q_2}{q_2^{\,0}}
\prod_{j_1=1}^n\frac{d^3k_{j_1}}{k_{j_1}} 
\prod_{j_2=1}^m\frac{d^3{k'}_{j_2}}{{k'}_{j_2}}
\nonumber\\
& &  \kern-2cm \times \int\frac{d^4y}{(2\pi)^4}
e^{iy\cdot(p_1+q_1-p_2-q_2-\sum k_{j_1}-\sum {k'}_{j_2})+ D_\rQCED}
\ \tilde{\bar\beta}_{n,m}(k_1,\ldots,k_n;k'_1,\ldots,k'_m),
\label{subp15b}
\end{eqnarray}}
where the new YFS-style~\cite{yfs1a} residuals
$\tilde{\bar\beta}_{n,m}(k_1,\ldots,k_n;k'_1,\ldots,k'_m)$ have $n$ hard gluons and $m$ hard photons and we illustrate the generic 2f 
final state with momenta $p_2,\; q_2$ specified for
definiteness. The infrared functions ${\rm SUM_{IR}(QCED)},\; D_\rQCED\; $
are defined in Refs.~\cite{qced1,irdglap1a}. Eq.\ (\ref{subp15b}) 
is exact to all orders in $\alpha$ and in $\alpha_s$.\par\indent 
The result Eq.\ (\ref{subp15b})
allows us to improve~\cite{irdglap1a} in the IR regime 
the DGLAP-CS~\cite{dglap1,cs1} kernels
as follows, using a standard notation:{\small
\begin{align}
P^{exp}_{qq}(z)&= C_F \FYFS(\gamma_q)e^{\frac{1}{2}\delta_q}\left[\frac{1+z^2}{1-z}(1-z)^{\gamma_q} -f_q(\gamma_q)\delta(1-z)\right],\nonumber\\
P^{exp}_{Gq}(z)&= C_F \FYFS(\gamma_q)e^{\frac{1}{2}\delta_q}\frac{1+(1-z)^2}{z} z^{\gamma_q},\nonumber\\
P^{exp}_{GG}(z)&= 2C_G \FYFS(\gamma_G)e^{\frac{1}{2}\delta_G}\{ \frac{1-z}{z}z^{\gamma_G}+\frac{z}{1-z}(1-z)^{\gamma_G}\nonumber\\
&\qquad +\frac{1}{2}(z^{1+\gamma_G}(1-z)+z(1-z)^{1+\gamma_G}) - f_G(\gamma_G) \delta(1-z)\},\nonumber\\
P^{exp}_{qG}(z)&= \FYFS(\gamma_G)e^{\frac{1}{2}\delta_G}\frac{1}{2}\{ z^2(1-z)^{\gamma_G}+(1-z)^2z^{\gamma_G}\},
\label{dglap19}
\end{align}}
where the superscript ``exp'' indicates that the kernel has been resummed as
predicted by Eq.\ (\ref{subp15b}) when it is restricted to QCD alone and 
where we refer the reader to Refs.~\cite{irdglap1a} for the
detailed definitions of the respective resummation functions $\FYFS,\gamma_A,\delta_A,f_G, A=q,G$
\footnote{The improvement in Eq.\ (\ref{dglap19}) 
should be distinguished from the 
resummation in parton density evolution for the ``$z\rightarrow 0$'' 
Regge regime -- see for example Ref.~\cite{ermlv,guido}. This
latter improvement must also be taken into account 
for precision LHC predictions.}.
See Refs.~\cite{qced1,irdglap1a}
for discussion of illustrative results and implications of the new 
kernels and Eq.\ (\ref{subp15b})
that are beyond the scope we have here. 
\par\indent
We have implemented the 
new IR-improved kernels in the HERWIG6.5 environment to produce
a new MC, HERWIRI1.0, which stands for ``high energy radiation with IR improvement''\footnote{We thank  M. Seymour and B. Webber for helpful
discussion on this point.}. We modify the kernels in the HERWIG6.5 module HWBRAN and in the attendant
 related modules\footnote{We thank M. Seymour and B. Webber for helpful discussion.} with the following substitutions:~{\small
$\text{DGLAP-CS}\; P_{AB}  \Rightarrow \text{IR-I DGLAP-CS}\; P^{exp}_{AB}$}
while leaving the hard processes alone for the moment. We have in 
progress~\cite{inprog}
the inclusion in our framework of YFS synthesized electroweak  
modules from Refs.~\cite{jad-ward1}
for HERWIG6.5, HERWIG++~\cite{herpp} and MC@NLO~\cite{mcnloa} hard processes\footnote{Similar results for PYTHIA~\cite{pythia} and for the new kernel evolution in Ref.~\cite{skrzjad1} are under study.}, as the
CTEQ~\cite{cteq1} and MRST(MSTW)~\cite{mrst1} 
best (after 2007) parton densities
do not include the precision electroweak higher order corrections that 
are needed in a 1\% precison tag budget for processes such as single 
heavy gauge boson production in the LHC environment~\cite{radcor-ew1}. 
\par
\begin{figure*}[t]
\centering
\setlength{\unitlength}{0.1mm}
\begin{picture}(1600, 830)
\put( 340, 750){\makebox(0,0)[cb]{\bf (a)} }
\put(1210, 750){\makebox(0,0)[cb]{\bf (b)} }
\put(   -70, 0){\makebox(0,0)[lb]{\includegraphics[width=80mm]{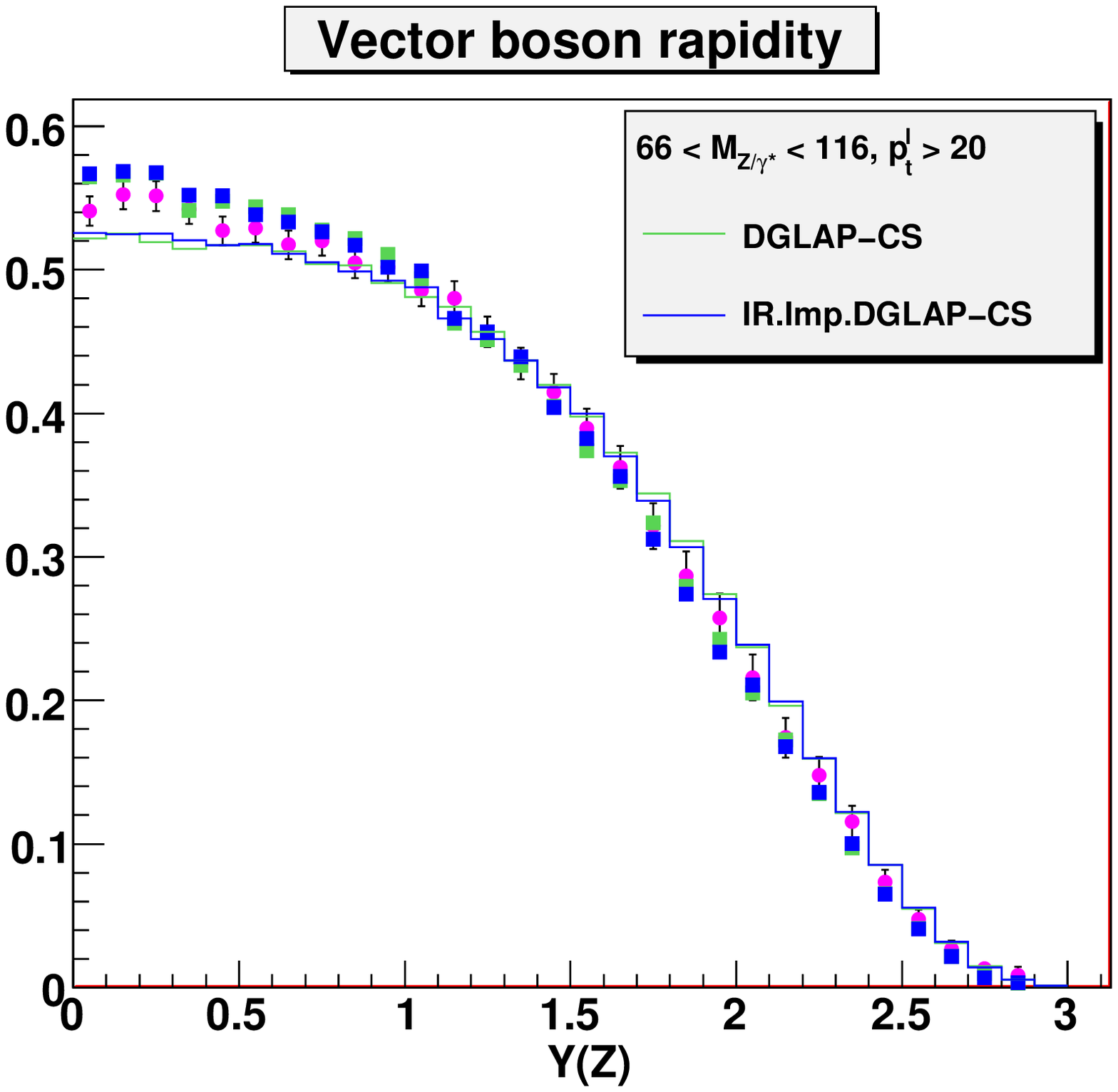}}}
\put( 800, 0){\makebox(0,0)[lb]{\includegraphics[width=80mm]{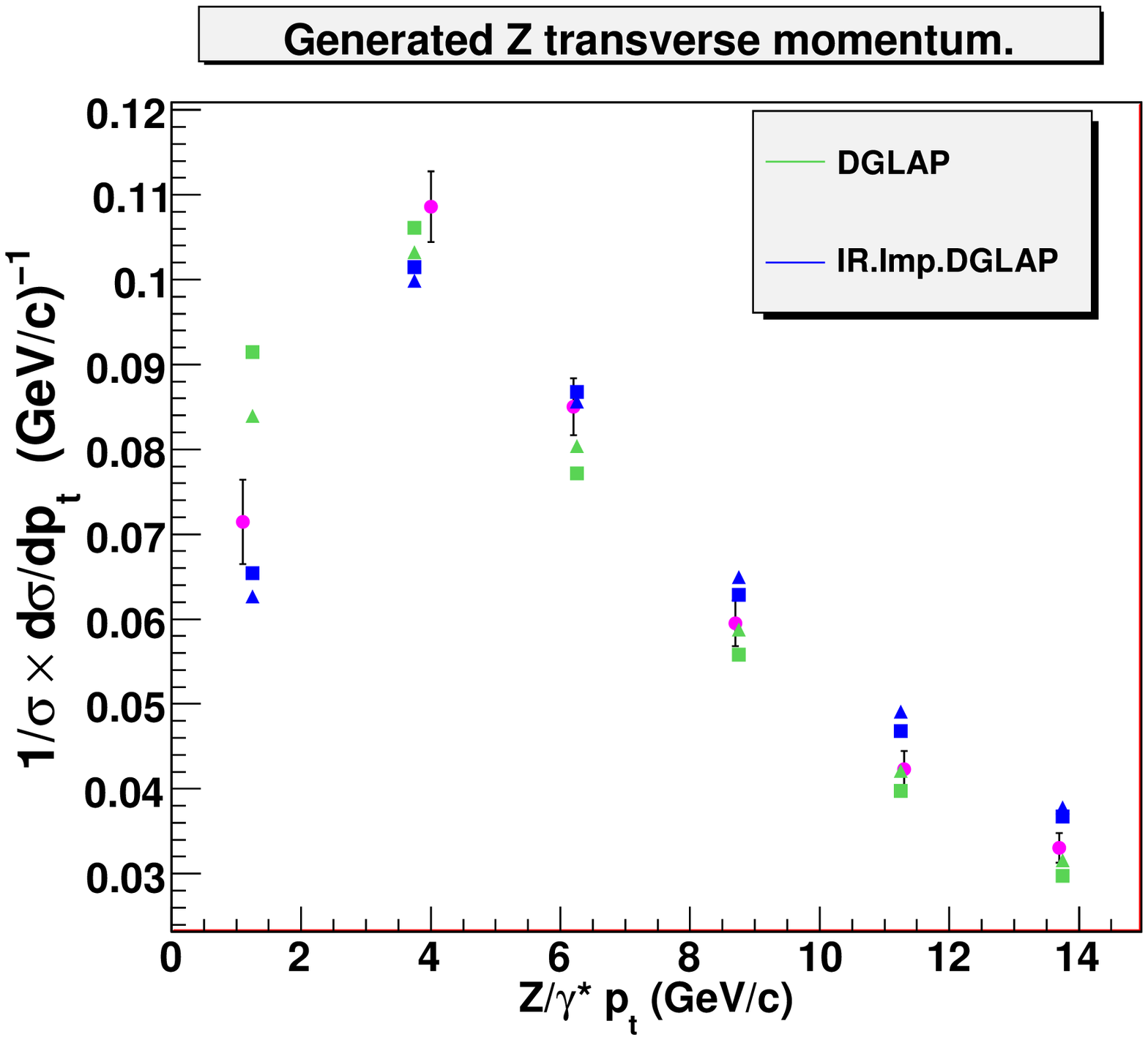}}}
\end{picture}
\caption{Comparison with FNAL data: (a), CDF rapidity data on
($Z/\gamma^*$) production to $e^+e^-$ pairs, the circular dots are the data, the green(blue) lines are HERWIG6.510(HERWIRI1.031); 
(b), D0 $p_T$ spectrum data on ($Z/\gamma^*$) production to $e^+e^-$ pairs,
the circular dots are the data, the blue triangles are HERWIRI1.031, the green triangles are HERWIG6.510 -- in both (a) and (b) the blue squares are MC@NLO/HERWIRI1.031, and the green squares are MC@NLO/HERWIG6.510. These are untuned theoretical results.
}
\label{fighw9}
\end{figure*}

\indent
The details of the implementation are given in Refs.~\cite{qced1,irdglap3-plb-prd} and we do not reproduce them here due to a lack
of space. We have done many comparisons of the properties of the
parton showers from HERWIG6.510 and HERWIRI1.031. In general, 
the IR-improved showers tend to be softer
in the energy fraction variable $z=E/E_{Beam}$ where $E\; (E_{Beam})$ is the cms parton(beam)
energy for hadron-hadron scattering respectively.
See Refs.~\cite{qced1,irdglap3-plb-prd} for the complete
discussion of such comparisons. We show in Fig.~\ref{fighw9}~\cite{irdglap3-plb-prd}
comparison 
analyses with the data from 
FNAL on the $Z$ rapidity and $p_T$ spectra as reported in 
Refs.~\cite{galea,d0pt}. We see that HERWIRI1.0(31)
and HERWIG6.5 both give a reasonable 
overall representation of the CDF rapidity data but that
HERWIRI1.031 is somewhat closer to the data for small values of $Y$(The $\chi^2$/d.o.f is 1.77(1.54) 
for HERWIG6.5 (HERWIRI1.0(31)).).  
Including the NLO
contributions to the hard process via MC@NLO/HERWIG6.510
and MC@NLO/HERWIRI1.031\cite{mcnloa}\footnote{We thank S. Frixione for
helpful discussions with this implementation.} improves the agreement for both
HERWIG6.510 and for HERWIRI1.031 (the $\chi^2$/d.o.f are changed
to 1.40 and 1.42 respectively).
That they are both consistent with
one another and within 10\% of the data in the low $Y$ region is
fully consistent with expectations and is an 
important cross-check on our work.  A more precise discussion at the NNLO
level with DGLAP-CS IR-improvement and
a more complete discussion  of the errors will
appear~\cite{elswh}.
We also see that HERWIRI1.031 gives a better fit to
the D0 $p_T$ data
compared to HERWIG6.510 for low $p_T$, 
(for $p_T<12.5$GeV, the $\chi^2$/d.o.f. are
$\sim$ 2.5 and 3.3 respectively if we add the statistical and systematic
errors), showing that the IR-improvement makes a better representation
of QCD in the soft 
regime for a given fixed order in perturbation theory.
Adding the ${\cal O}(\alpha_s)$ correction
from MC@NLO~\cite{mcnloa}
improves the $\chi^2$/d.o.f for
the HERWIRI1.031 in both the soft and hard regimes and it improves
the HERWIG6.510 $\chi^2$/d.o.f for $p_T$ near $3.75$ GeV
where the distribution peaks. For $p_T<7.5$GeV the $\chi^2$/d.o.f for
the MC@NLO/HERWIRI1.031 is 1.5 whereas that for MC@NLO/HERWIG6.510 is 
worse. We await further tests of the new
approach, both at FNAL and at LHC. -- One of us (B.F.L.W) acknowledges helpful discussions with Prof. Bryan Webber
and Prof. M. Seymour and with Prof. S. Frixione. B.F.L. Ward also thanks Prof. L. Alvarez-Gaume and Prof. W. Hollik for the support and kind hospitality of the CERN TH Division and of the Werner-Heisenberg Institut, MPI, Munich, respectively, while this work was in progress. S. Yost acknowledges the hospitality 
and support of Princeton University and the CERN TH Division.




\begin{thebibliography}{99}
\bibitem{qedeffects1} S. Haywood, P.R. Hobson, W. Hollik and Z. Kunszt,
in {\it Proc. 1999 CERN Workshop on Standard Model Physics ( and more )
at the LHC, CERN-2000-004}, eds. G. Altarelli and M.L. Mangano,( CERN,
Geneva, 2000 ) p. 122;
 H. Spiesberger, {\it Phys. Rev.} D{\bf 52} ( 1995 ) 4936;
 W.J. Stirling,''Electroweak Effects in Parton Distribution 
Functions'', talk presented at ESF Exploratory Workshop,
{\it Electroweak Radiative Corrections to Hadronic Observables at TeV
Energies }, Durham, Sept., 2003;
 M. Roth and S. Weinzierl,{\it Phys. Lett.} {\bf B590} (2004) 190;
 J. Blumlein and H. Kawamura, {\it Nucl. Phys.} {\bf B708} (2005) 467; {\it Acta Phys. Pol.} {\bf B33} (2002) 3719; 
 W. J. Stirling {\it et al.}, 
in {\it Proc. ICHEP04}, eds. H. Chen {\it et al.}
(World Sci. Publ., Singapore, 2005) p. 527;
 A.D. Martin {\it et al.}, {\it Eur. Phys. J.} {\bf C39} (2005) 155, and references therein. 
\bibitem{radcor-ew1} A. Kulesza {\it et al.}, {\it PoS RADCOR2007}: 001, 2007;
 A. Denner {\it et al.}, {\it PoS RADCOR2007}: 002, 2007;
 {\it  Nucl.Phys.} {\bf B662} (2003) 299;
 G. Balossini {\it et al.}, arXiv:0805.1129; S. Dittmaier, in {\it Proc. LP09}, 2009, in press,
and references therein.
\bibitem{jadach1a} See for example S. Jadach {\it et al.}, in {\it Geneva 1995, Physics at LEP2, vol. 2}, pp. 229-298; hep-ph/9602393, for a discussion of technical and physical precision.
\bibitem{qced1} C. Glosser, S. Jadach, B.F.L. Ward and S.A. Yost,{\it Mod. Phys. Lett. A}{\bf 19}(2004) 2113;
 B.F.L. Ward, C. Glosser, S. Jadach and S.A. Yost, in {\it Proc. ICHEP04, vol. 1}, eds. H. Chen {\it et al.},(World. Sci. Publ. Co., Singapore, 2005) p. 588; B.F.L. Ward and S. Yost, {\it Acta Phys. Polon.} {\bf B38} (2007) 2395; {\it PoS RADCOR2007}: 038, 2007; B.F.L. Ward {\it et al.}, in {\it Proc. ICHEP08}, Philadelphia, 2008, eConf C080730, [arXiv:0810.0723]; in {\it Proc. 2008 HERA-LHC Workshop},DESY-PROC-2009-02, eds. H. Jung and A. De Roeck, (DESY, Hamburg, 2009)pp. 180-186, and references therein.
\bibitem{herwig1} G. Corcella {\it et al.}, preprint hep-ph/0210213; 
{\it J. High Energy Phys.} {\bf 0101} (2001) 010;
 G. Marchesini {\it et al.}, {\it Comput. Phys. Commun.}{\bf 67} (1992) 465.
\bibitem{irdglap3-plb-prd} S. Joseph {\it et al.}, Phys. Lett. B{\bf 685}, 283 (2010); Phys. Rev. D{\bf 81}, 076008 (2010).
\bibitem{irdglap1a} B.F.L. Ward, {\it Adv. High Energy Phys.} {\bf 2008} (2008) 682312; {\it Ann. Phys.} {\bf 323} (2008) 2147.
\bibitem{cattrent1} G. Sterman,{\it Nucl. Phys.} {\bf B281}, 310 (1987); S. Catani and L. Trentadue,
{\it Nucl. Phys.} {\bf B327}, 323 (1989); {\it ibid.} {\bf B353}, 183 (1991).
\bibitem{scet1} See for example C. W. Bauer, A.V. Manohar and M.B. Wise, {\it Phys. Rev. Lett.} {\bf 91} (2003) 122001; {\it Phys. Rev.} {\bf D70} (2004) 034014.
\bibitem{dglap1}
G. Altarelli and G. Parisi, {\it Nucl. Phys.} {\bf B126} (1977) 
298;
 Yu. L. Dokshitzer, {\it Sov. Phys. JETP} {\bf 46} (1977) 641;
 L.~N. Lipatov, {\it Yad. Fiz.} {\bf 20} (1974) 181;
 V. Gribov and L. Lipatov,
{\it Sov. J. Nucl. Phys.} {\bf 15} (1972) 675;
 V. Gribov and L. Lipatov,
{\it Sov. J. Nucl. Phys.} {\bf 15} (1972) 938;
 see also J.C. Collins and J. Qiu,
{\it Phys. Rev. D}{\bf 39} (1989) 1398 for an alternative discussion
of DGLAP-CS theory.
\bibitem{cs1}C.G. Callan, Jr., {\it Phys. Rev. D}{\bf 2} (1970) 1541; K. Symanzik, 
{\it Commun. Math. Phys.} {\bf 18} (1970) 227; K. Symanzik, in {\em Springer Tracts in Modern Physics},
{\bf 57}, ed. G. Hoehler (Springer, Berlin, 1971) p. 222; see also
S. Weinberg, {\it Phys. Rev. D}{\bf 8} (1973) 3497. 
\bibitem{scott1a} N.E. Adam {\it et al.}, {\it J. High Energy Phys.} {\bf 05} (2008) 062; {\it ibid.} {\bf 09} (2008) 133; {\it ibid.} {\bf 11} (2010) 074.
\bibitem{yfs1a} D.~R.~Yennie, S.~C.~Frautschi, and H.~Suura, {\it Ann. Phys.} {\bf 13} (1961) 379; see also K.~T.~Mahanthappa, {\it Phys.~Rev.}~{\bf 126} (1962) 329.
\bibitem{ermlv} B.I. Ermolaev, M. Greco and S.I. Troyan, {\it PoS DIFF2006} (2006) 036, and references therein.
\bibitem{guido} G. Altarelli, R.D. Ball and S. Forte, 
{\it PoS RADCOR2007} (2007) 028.
\bibitem{inprog} M. Hejna {\it et al.}, to appear.
\bibitem{jad-ward1} S. Jadach and B.F.L. Ward, {\it Comput. Phys. 
Commun.} {\bf 56}(1990) 351; {\it Phys.Lett.} {\bf B274} (1992) 470; S. Jadach {\it et al.}, {\it Comput. Phys. Commun.} {\bf 102}
(1997) 229; S. Jadach, B.F.L. Ward and Z. Was, {\it Phys. Rev. D} {\bf 63} (2001) 113009; {\it Comp. Phys. Commun.} {\bf 130} (2000) 260; S. Jadach {\it et al.}, {\it ibid.}{\bf 140} (2001) 432, 475, and references therein.
\bibitem{herpp} M. Bahr {\it et al.}, arXiv:0812.0529 and references therein.
\bibitem{mcnloa} S. Frixione and B.Webber, {\it J. High Energy Phys.} {\bf 0206} (2002) 029; S. Frixione, P. Nason and B. Webber, {\it ibid.} {\bf 0308} (2003) 007.
\bibitem{pythia} T. Sjostrand {\it et al.}, hep-ph/0308153.
\bibitem{skrzjad1} S. Jadach and M. Skrzypek, 
{\it Comput. Phys. Commun.} {\bf 175} (2006) 511; P. Stevens {\it et al.}, {\it Acta Phys. Polon.} {\bf B38} (2007) 2379, and references therein.
\bibitem{cteq1} F. Olness, private communication; P.M. Nadolsky {\it et al.}, arXiv:0802.0007.
\bibitem{mrst1} R. Thorne, private communication; A.D. Martin {\it et al.}, arXiv:0901.0002 and references therein.
\bibitem{galea} C. Galea, in {\it Proc. DIS 2008}, London, 2008,\newline 
\verb$http://dx.doi.org/10.3360/dis.2008.55$.
\bibitem{d0pt} V.M. Abasov {\it et al.}, {\it Phys. Rev. Lett.} {\bf 100}, 102002 (2008).
\bibitem{elswh} B.F.L. Ward {\it et al.}, to appear.
\end{thebibliography}
\end{document}